\documentclass[overric,runningheads]{llncs}

\usepackage{float}
\restylefloat{table}
\usepackage{graphicx}
\usepackage{cite}
\usepackage{pgfplots, pgfplotstable}
\usepackage{tikz}
\usepackage{amssymb}
\usepackage{amsmath}
\usepackage{subcaption}
\usepackage{xcolor,listings}
\usepackage[misc]{ifsym}
\usepackage[ruled,lined,linesnumbered,vlined]{algorithm2e}
\usepackage[hyphenbreaks]{breakurl}
\usepackage[hyphens]{url}
\usepackage{hyperref}
\usepackage{algpseudocode}
\usepackage{siunitx}
\usepackage{wrapfig}
\usetikzlibrary{decorations.text}

\newcommand{\cas}{\mathcal{C}}
\newcommand{\attr}{\mathcal{A}}
\newcommand{\ev}{\mathcal{E}}

\newcommand{\dfr}{\textgreater_L}

\titlerunning{Native Directly Follows Operator}

\title{Native Directly Follows Operator}

\author{Alifah Syamsiyah$^{(\textrm{\Letter})}$, Boudewijn F. van Dongen, Remco M. Dijkman}
\authorrunning{Alifah Syamsiyah, Boudewijn F. van Dongen, Remco M. Dijkman}

\institute{Eindhoven University of Technology\\
	\email{A.Syamsiyah@tue.nl, B.F.v.Dongen@tue.nl, R.M.Dijkman@tue.nl}}

\begin{document}
	
\maketitle

\begin{abstract}

Typical legacy information systems store data in relational databases.
Process mining is a research discipline that analyzes this data to obtain insights into processes. Many different process mining techniques can be applied to data.
In current techniques, an XES event log serves as a basis for analysis.
However, because of the static characteristic of an XES event log, we need to create one XES file for each process mining question, which leads to overhead and inflexibility.
As an alternative, people attempt to perform process mining directly on the data source using so-called intermediate structures.
In previous work, we investigated methods to build intermediate structures on source data by executing a basic SQL query on the database.
However, the nested form in the SQL query can cause performance issues on the database side.
Therefore, in this paper, we propose a native SQL operator for direct process discovery on relational databases.
We define a native operator for the simplest form of the intermediate structure, called the ``directly follows relation''.
This approach has been evaluated with big event data and the experimental results show that it performs faster than the state-of-the-art of database approaches. 
\end{abstract}


\section{Introduction}
\label{sec:intro}

Process mining is a research discipline that turns event data into process models, checks the model with reality, and enhances the model with statistics derived from event data.
There has been an extensive research in process mining, including \emph{process discovery, conformance checking}, and \emph{enhancement}.

Typical legacy information systems store data in relational databases.
In the context of big data, the stored data possesses the characteristic of 4V (Volume, Variety, Veracity, and Velocity). When applying process mining, a process analyst aims to find process structures in the event data.
In some cases, however, the process analyst simply does not know what kind of process view can be obtained from the data. In this scenario, many process mining  questions are to be answered on the underlying data in order to get insights into the business processes. These questions may involve various classifiers, span on multiple times, and use heterogeneous case notions.

Let us take a process in a commercial bank as an example.
The financial manager may want to figure out the core process of the bank, such as ``the process involving successful loans during last month'', ``the progress of ongoing investment projects with a property this year'', ``how is the performance of collaboration between department X and Y'', and many other process mining related questions.

In the current state-of-the-art, process mining tools need (XES) event logs as  input.
One of the characteristics of event logs is that they are static, i.e. one file only contains one case notion and contains data for one specific period of time.
Therefore, we need to create one log file for each process mining question.
This produces unnecessary overhead caused by the exporting from the data source, conversion to a tool's input format, and importing into the process mining tool.
Moreover, the traditional techniques lack of flexibility, i.e. once we change the perspective from which we look at the data, we need to change the log file.

Given the fact that many legacy information systems use a database as the back end, we study the question how to perform a process mining directly on the database \cite{dbxes-ext,bpi2015,recurrent,bpmds2017}.
In this paper, we focus on process discovery, i.e. we focus on deriving a process model from event data. In a process discovery algorithm, there is typically a so called \emph{intermediate structure}, which is a first abstraction of the event data.
In \cite{dbxes-ext}, we showed how to compute this intermediate structure inside a database and how to only import the structure (not the event data) into a process mining tool. The mining tool then discovers the process model using an existing algorithm.

In order to compute the intermediate structure, the existing approach uses standardized SQL queries, which, unfortunately, are not designed towards process mining purposes.
Let us take a simple intermediate structure called \emph{Directly Follows Relation} (DFR).
To compute DFR, the standard SQL query has a nested form which results in a bad performance. In this paper we present two ways to overcome nested queries and we compare the performance of all database  variants against the traditional approach.

A way to overcome the nested queries is to create an interface between a database and a process mining tool. 
As computing the DFR is based on a sorted event data, through this interface, we execute a standard SQL query to sort event data in a database.
Then, we import the sorted event data into a process mining tool and compute the frequency of the DFR in the tool.
We later show in this paper that even though this approach does not contain a nested structure, it still has a performance issue as the whole log needs to be transferred to the process mining tool.

In \cite{remco} it has been proven that executing a nested query to compute the DFR leads to third order polynomial time complexity if the intermediate results of the queries do not fit into memory anymore. To overcome this problem, a native SQL operator is proposed.

In this paper, we propose a native SQL operator for direct process discovery in relational databases.
The native operator is designed for a specific process discovery purpose.
As a starting point, this paper investigates the directly follows relation.
However, it does not restrict the possibility to extend the operator to other kinds of intermediate structures.
Using this native operator, the database has more flexibility to query process mining related questions.
Moreover, it harnesses the database technology to speed up the computation time.





The reminder of this paper is structured as follows.
Section \ref{sec:preliminaries} introduces some important terminologies that are used in the paper.
Then we introduce the proposed idea, which is the native SQL operator for a direct process discovery, in Section \ref{sec:native}.
The implementation of this operator is given in Section \ref{sec:impl} and Section \ref{sec:experiment} demonstrates the experimental results.
Finally the paper is concluded in Section \ref{sec:concl}.

\section{Preliminaries}
\label{sec:preliminaries}

This paper deals with process mining.
We refer to Process Mining Manifesto \cite{manifesto} and Process Mining book in \cite{process-mining-book-2} for more detail explanation about process mining.
Process mining needs event data as inputs.
Events are collections of attributes, referring to the executions of activities in the context of some cases in a process.

\begin{definition} [Event Attribute]
	Let $\attr$ be the universe of event attributes, $\cas$ be the universe of cases, $\ev$ be the universe of events,
	and $E \subseteq \ev$ be a collection of events. 
	
	For any event $e \in E$ and name $a \in \attr$: $\#_a(e)$ is the value of attribute $a$ for event $e$, $\#_a(e) = \bot$ if there is no value. 
	$\{$caseid, act, time$\}\in \attr$ are standard event attributes, such that $\#_{caseid}(e)$ is the case of event $e$, $\#_{act}(e)$ is the activity name of event $e$, and $\#_{time}(e)$ is the timestamp when event $e$ is executed.

\end{definition}


	


Furthermore, an event log is a collection of events captured within a particular time period.

\begin{definition}[Event Log]	
	Let $E \subseteq \ev$ be a collection of events and $t_s,t_e \in \mathbb{R}$ two timestamps with $t_s<t_e$ relating to the start and the end of the collection period.

	A case $\sigma \in E^*$ is a sequence of events such that the same event occurs only once in $\sigma$, i.e. $|\sigma| = |\{ e \in \sigma \}|$. Furthermore, each event in a case refers to the same case $c \in \cas$, i.e. $\forall_{e \in \sigma}\#_{case}(e)= c$ and we assume all events within the given time period are included, i.e. $\forall_{e \in \ev}  (\#_{case}(e)= c \wedge t_s \leq \#_{time}(e) \leq t_e) \implies e \in \sigma$.
	
	An event log $L \subseteq E^*$ is a set of cases.		
\end{definition}

%
%
%

Finally, we define Directly Follows Relation (DFR), which holds for two activities $A$ and $B$ if and only if somewhere in the event log $L$, there are two successive events in a trace corresponding to these activities.
Note that we also consider cases with partial ordered events.

\begin{definition}[Directly Follows Relation (DFR)]
	\label{def:directSuccession}\\
	Let $\attr$ be the universe of event attributes, let $\ev$ be a universe of events and $L$ be an event log over $E \subseteq \ev$.
	Let $\mathcal{M} = \{ a \in \attr\ | e \in E \wedge \#_{act}(e) =a  \}$ be the set of activities in the log. 
	The DFR $\textgreater_L: \mathcal{M} \times \mathcal{M} \rightarrow \mathbb{N}$ counts the number of times activity $a$ is directly followed by activity $b$ in some cases in $L$ as follows: \\
	$\textgreater_L(a,b) = \Sigma_{\sigma \in L} \ \Sigma_{i = 1}^{|\sigma|-1}
	\begin{cases}
	1, &\emph{if} \ i < j \ \wedge \#_{act}(\sigma(i)) = a \ \wedge \ \#_{act}(\sigma(j)) = b \ \wedge  \\
	  &\#_{time}(\sigma(i)) < \#_{time}(\sigma(j)) \ \wedge \\
	  &\neg\exists k \ \#_{time}(\sigma(i)) < \#_{time}(\sigma(k)) < \#_{time}(\sigma(j)) \\		
	0, &\emph{otherwise}.				
	\end{cases}$\\
\end{definition}

\section{Native Directly Follows Operator}
\label{sec:native}

In this section, we introduce a new \emph{native directly follows operator} for computing directly follows relation.
In the following, we first explain the input and output of this operator, an example of how to use the operator, and how to process the result in the context of discovery. 

\begin{figure*}[!t]
	\centering
	\includegraphics[width=0.8\textwidth]{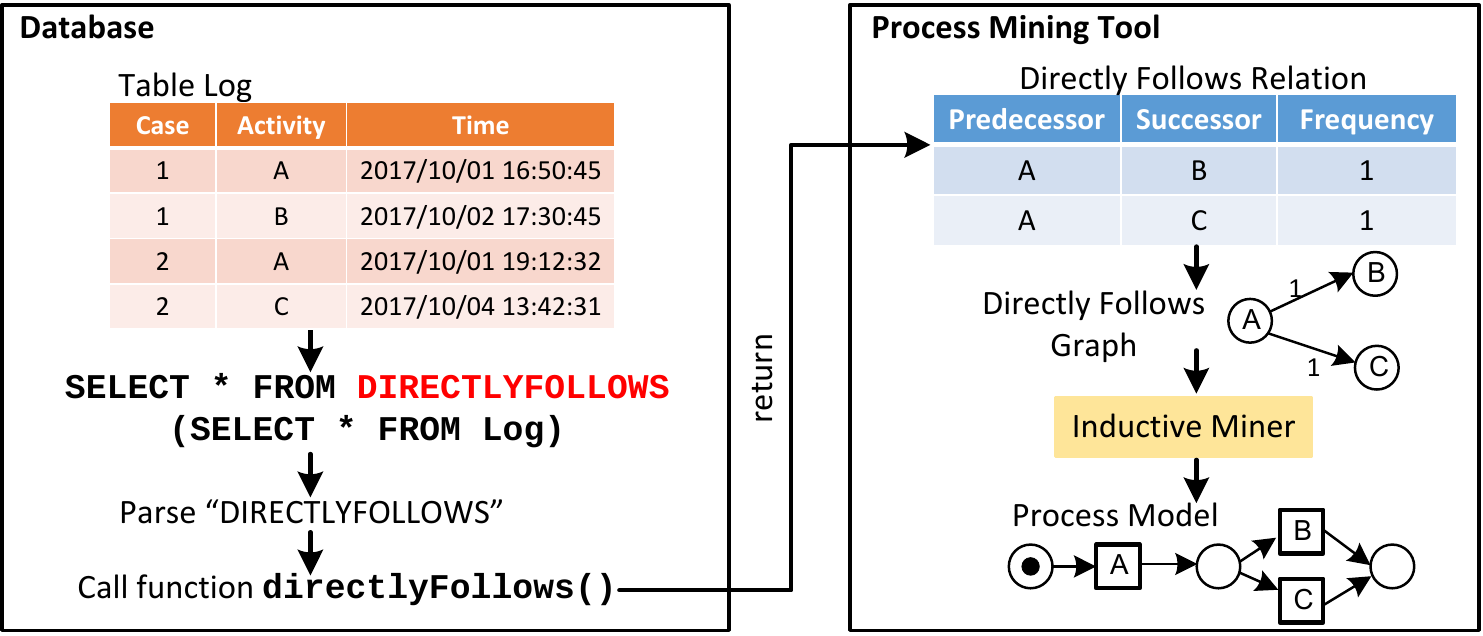}
	\caption{Using native operator ``directlyfollows'' and Inductive Miner to discover a process model}
	\label{fig:native}
\end{figure*}

A native operator \texttt{directlyfollows} requires a table object representing the event log, which consists of three columns: the case, the activity, and  the timestamp.
As a result, it returns a table object representing the DFR, which consists of three columns: the first and the second part of pair of the DFR, and the frequency of each pair.
The schema in Figure \ref{fig:native} illustrates how to utilize the operator and what happens inside a database engine and a process mining tool assuming that we use Inductive Miner for the discovery.

\begin{table}[!bt]\scriptsize
	\caption{(a) Example of event data stored in Table $Log$, (b) The DFR of Table $Log$}
	\begin{subtable}{.5\linewidth}
		\centering
		\caption{}
		\begin{tabular}{|l|l|l|}
			\hline
			Case & Activity & Time \\
			\hline
			1 & Send request & 2017/10/01\\
			1 & Check application & 2017/10/02\\
			1 & Check document & 2017/10/02\\
			1 & Accept & 2017/10/05\\
			2 & Send request & 2017/10/03 \\
			2 & Check application & 2017/10/07\\
			2 & Reject & 2017/10/10\\
			\hline
		\end{tabular}
	\end{subtable}%
	\begin{subtable}{.5\linewidth}
		\centering
		\caption{}
		\begin{tabular}{|l|l|l|}
			\hline
			Event$\_$Label$\_$P & Event$\_$Label$\_$S & Frequency \\
			\hline
			Send request & Check application & 2 \\
			Send request & Check document & 1 \\
			Check application & Accept & 1 \\
			Check document & Accept & 1 \\
			Check application & Reject & 1 \\
			\hline
		\end{tabular}
	\end{subtable} 
	\label{tab:log}
\end{table}

Assume that we have a Table $Log$ (see Table \ref{tab:log}a), we compute the DFR using the native operator \texttt{directlyfollows} as follows.

\begin{lstlisting}[
language=SQL,
showspaces=false,
basicstyle=\ttfamily,
numbers=left,
numberstyle=\tiny,
commentstyle=\color{gray}
]
SELECT * FROM DIRECTLYFOLLOWS (SELECT * FROM Log);
\end{lstlisting}

\begin{algorithm}[!h]
	\caption{$\texttt{directlyFollows}$}
	\label{algo:dfr}
	
	\KwIn{$L$}
	\KwOut{$\dfr$}
	
	

	\ForEach{$\sigma \in L$}
	{
		sort($\sigma$) \\ 
		$sa \leftarrow 0$ \\
		$ea \leftarrow 0$ \\
		$aTime \leftarrow \#_{time}(\sigma(sa))$ \\
		\While{$ea + 1 < |\sigma|$ \emph{and} $aTime = \#_{time}(\sigma(ea + 1))$ }
		{
			$ea \leftarrow ea + 1$
		}
		$sc \leftarrow ea + 1$ \\
		$ec \leftarrow sc$ \\
		\While{$ec < |\sigma|$}
		{
			$cTime \leftarrow \#_{time}(\sigma(sc))$ \\
			\While{$ec + 1 < |\sigma|$ \emph{and} $cTime = \#_{time}(\sigma(ec + 1))$ }
			{
				$ec \leftarrow ec + 1$
			}
			\For{$i \leftarrow sa; i \le ea; i++$}
			{
				\For{$j \leftarrow sc; j \le ec; j++$} 
				{
					$a \leftarrow \#_{act}(\sigma(i))$ \\
					$b \leftarrow \#_{act}(\sigma(j))$ \\
					$freq \leftarrow \dfr(a, b)$ \\
					$freq \leftarrow freq + 1$ \\
					$\dfr(a, b) \leftarrow freq$
				}	
			}
			$sa \leftarrow sc$ \\
			$ea \leftarrow ec$ \\
			$sc \leftarrow ea + 1$ \\
			$ec \leftarrow sc$
		}
	}

	\Return{$\dfr$}

\end{algorithm}

After the database engine parses \texttt{directlyfollows}, it calls the \texttt{directly} \texttt{Follows()} function (see Algorithm \ref{algo:dfr}).
First of all, for each case $\sigma$ in log $L$, the function sorts $\sigma$.
Then it assigns initial value for four variables, namely (1) $sa$ (start antecedent), (2) $ea$ (end antecedent), (3) $sc$ (start consequent), and (4) $ec$ (end consequent).
These four variables point to indexes in $\sigma$ such that all events between the start and end indexes have the same timestamp, i.e. 
events between $sa$ and $ea$ have the same timestamp, events between $sc$ and $ec$ have the same timestamp.
Based on these indexes, the function creates combinations between antecedent events and consequent events to construct a pair of directly follows relation.
Finally, it counts the frequency of such pair and returns the directly follows (as denoted in Table \ref{tab:log}b).

In Java, we use hash table to store the relations. Pairs are the key and frequencies are the value.
The complexity is worst case $|E|$ log $|E|$ (with $E$ is the total number of events) due to sorting events.

In the next step, the DFR is retrieved by a process mining tool.
Since we use the Inductive Miner for the discovery, such relation is converted into a directly follows graph before the algorithm constructs a process model.
Note that the result from the \texttt{directlyfollows} operator can be used directly  in the existing process discovery technique without modifying the algorithm or reinvent a new discovery method.

Process discovery using the native operator has several advantages. 
\emph{First}, the query can be expressed in a straightforward way, hence it is more convenient for novice users to express various process mining related questions.
\emph{Second}, the query does not contain a nested form which leads to bad performance.
\emph{Third}, the computation (i.e. the abstraction phase) can be done inside relational databases, thus leveraging the power of database technology and saving the memory usage of process mining tool.
\emph{Fourth}, there is no need to extract and load a log file into a process mining tool, thus saving time and reducing the complexity.
\emph{Fifth}, the DFR can be computed upon insertion of data using the standard triggering mechanism of any database system, hence eliminating the need for the process analyst to wait for the computation to finish.

\section{Implementation}
\label{sec:impl}

We implemented the \texttt{directlyfollows} as a native operator in H2 Database\footnote{\url{https://github.com/alifahsyamsiyah/h2processmining-master}}.
Moreover, process discovery using this operator and the Inductive Miner were implemented as a plug-in in an open source process mining toolkit called ProM.
The plug-in name is \emph{H2 Inductive Miner} and it is distributed within the \emph{DatabaseInductiveMiner} package\footnote{\url{https://svn.win.tue.nl/repos/prom/Packages/DatabaseInductiveMiner/Trunk/}}.

\begin{figure}[!h]
	\centering
	\begin{subfigure}{.99\textwidth}
		\includegraphics[width=1\linewidth]{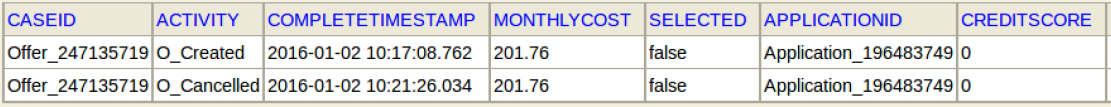}
		\label{fig:subdb1}
	\end{subfigure}
	\qquad
	\begin{subfigure}{.99\textwidth}
		\includegraphics[width=1\linewidth]{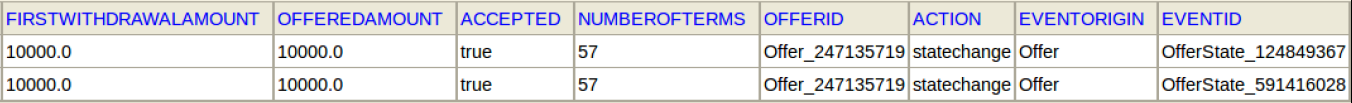}
		\label{fig:subdb2}
	\end{subfigure}
	\caption{The excerpt of Table BPI2017 in H2 database}
	\label{fig:db}
\end{figure}

Figure \ref{fig:db} illustrates the excerpt of the log in the server.

\begin{figure}[!h]
	\centering
	\begin{minipage}{.4\textwidth}
		\centering
		\includegraphics[width=0.9\textwidth]{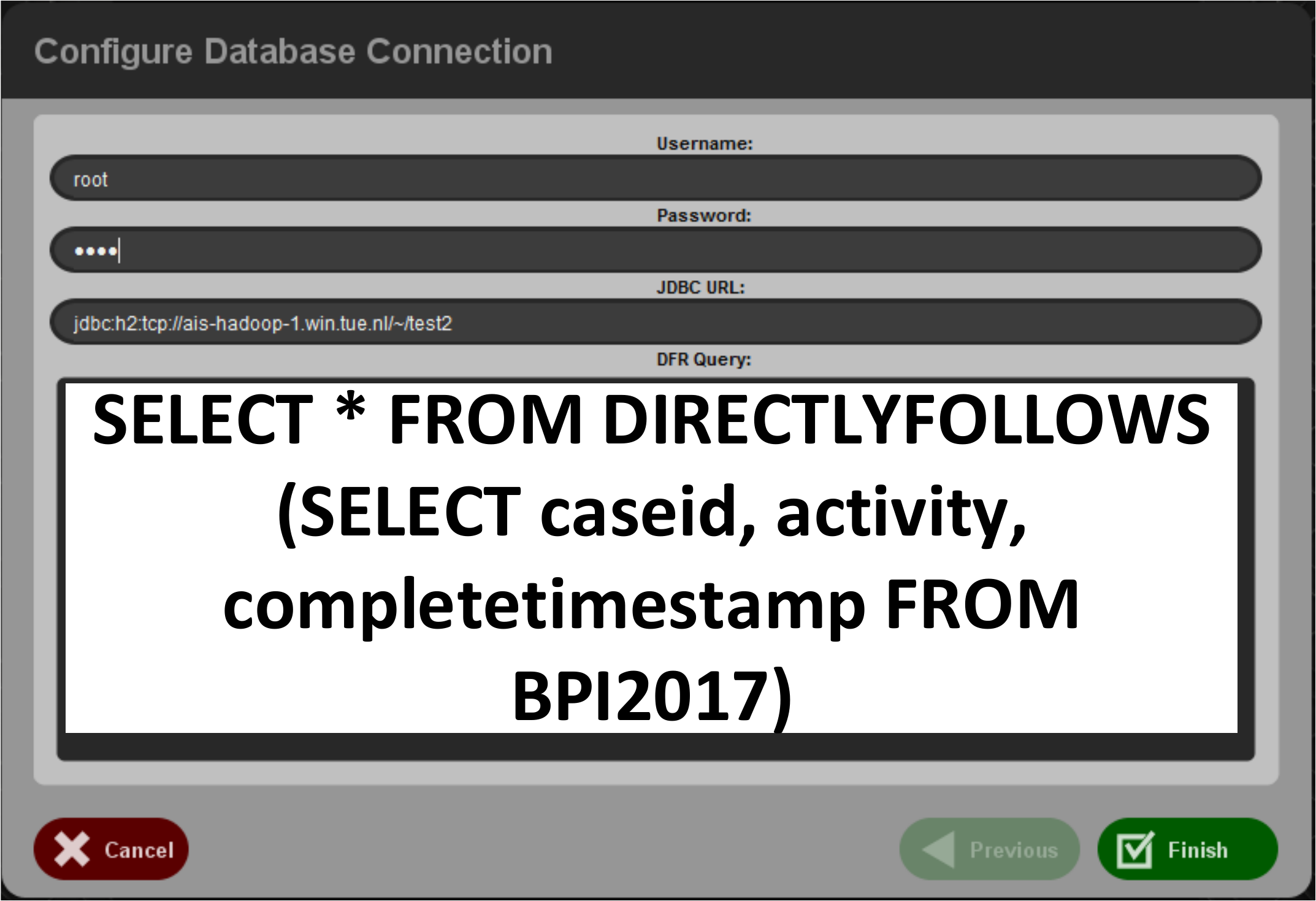}
		\caption{A query to extract the DFR from Table BPI2017}
		\label{fig:query}
	\end{minipage}%
	\begin{minipage}{.6\textwidth}
		\centering
		\includegraphics[width=0.99\textwidth]{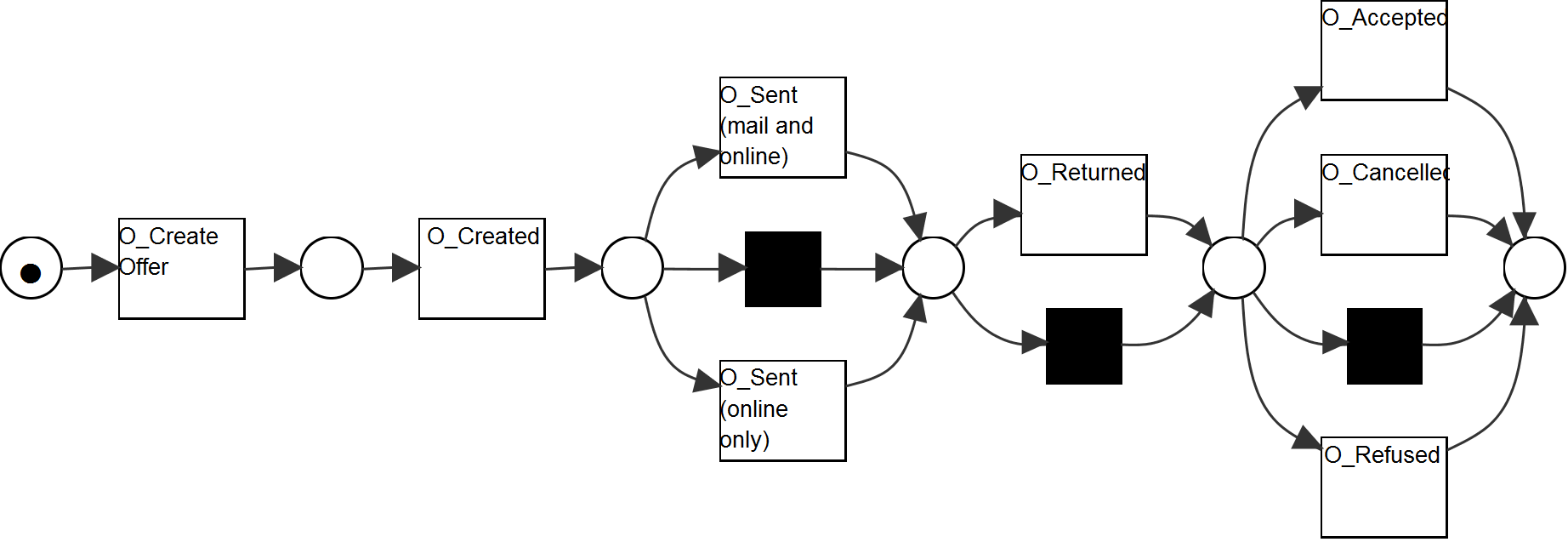}
		\caption{The discovered model of Table BPI2017}
		\label{fig:model}
	\end{minipage}
\end{figure}

As an example, we used the offer event log of BPI Challenge 2017 \cite{BPI17}, which was imported into an H2 database server.
To connect with the server, we set the required database configuration in the \emph{H2 Inductive Miner} plug-in.
We filled in the username and the password of the database, the JDBC URL, and the query to extract the DFR as denoted in Figure \ref{fig:query}. 
After we clicked the ``Finish'' button, the plug-in automatically executes the query, retrieved the DFR, converts the values into a directly follows graph, and finally discovers a process model based on the Inductive Miner algorithm.
The result is displayed in Figure \ref{fig:model}.

\section{Experimental Results}
\label{sec:experiment}

\begin{figure*}[!t]
	\centering
	\includegraphics[width=0.5\textwidth]{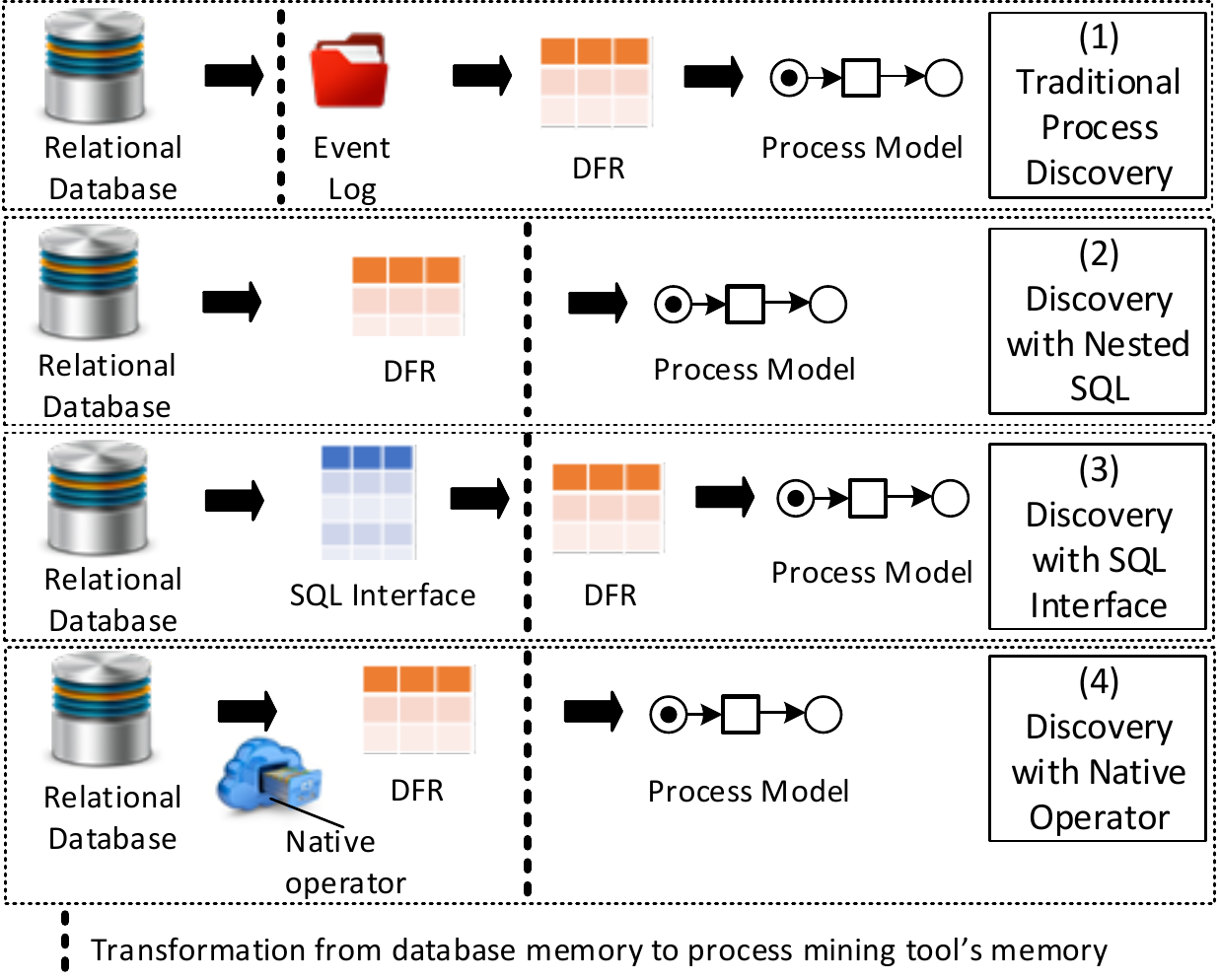}
	\caption{Four different approaches in process discovery: (a) traditional process discovery, (b) discovery with Nested SQL, (c) discovery with SQL interface, and  (d) discovery with native operator}
	\label{fig:app}
\end{figure*}

We implemented our work in ProM and H2 and we compared 
the native operator with the other three approaches  (Figure \ref{fig:app}): (1) the traditional technique, (2) the discovery with Nested SQL, and (3) the SQL interface. The starting point for the traditional approach is an XES event log file already loaded into memory. For the database approaches, the starting point is a database in which events have been inserted.
In the initial experiments, we did not use any type of indices for the events so we can represent general scenarios where table logs are constructed without any indices.
However, in the subsequent experiments, we also investigate scenarios where indices are utilized.
We used an H2 database server which has 64GB of RAM and 8 cores of CPU @2.40Ghz.
Furthermore, the discovery was executed in a personal computer which has 8GB of RAM and 2 cores of CPU @2.30Ghz.

We created synthetic logs with the number of events ranging from 1K to 42M with between 30 to 3840 activities.
More specifically, we created two kinds of synthetic logs: (a) logs with an increased number of activities, and (b) logs with an increased number of events.
For (a), we relabeled activities to extend the number of activities while keeping the same number of events.
For (b), we merged one case with another case to extend the number of events while keeping the same number of activities.
This way, we preserve the control flow for our extended logs in the same way as the control flow of the original log.
%

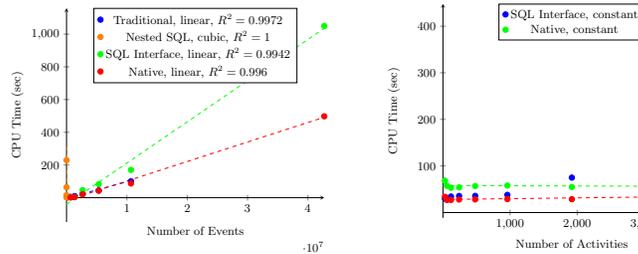
\begin{figure}[!t]
	\begin{center}
		
		\begin{tikzpicture}[scale=0.5]

		\begin{axis}[
		axis x line=middle,
		axis y line=middle,
		enlarge y limits=true,
		ylabel= CPU Time (sec),
		xlabel= {Number of Events},
		x label style={at={(axis description cs:0.5,0)},anchor=north},
		y label style={at={(axis description
		 cs:-0.15,.5)},rotate=90,anchor=south},
		legend style={at={(0.5,1)}, anchor=north}		
		]

		\addplot[only marks, blue] table  {result/abs-trad-event.dat};

		\addplot[only marks, orange] table  {result/abs-nested-event.dat};
		
		\addplot[only marks, green] table  {result/abs-interface-event.dat};
		
		\addplot[only marks, red] table  {result/abs-native-event.dat};
		
		\addplot+[dashed,mark=none,
					samples=10,color=blue,domain=0:10677776] 
					{1e-05*x - 2.7183};
					
		\addlegendentry{Traditional, linear, $R^2 = 0.9972$}	
		
		\addplot+[dashed,mark=none,
					samples=10,color=orange,domain=0:13792] 
					{-2e-11*x*x*x + 2e-06*x*x - 0.0014*x + 0.9362};
		
		\addlegendentry{Nested SQL, cubic, $R^2= 1$}	

		\addplot+[dashed,mark=none,
					samples=10,color=green,domain=0:42711104] 
					{2.519204e-05*x -4.015187e+01};
					
		\addlegendentry{SQL Interface, linear, $R^2= 0.9942$}	
		
		\addplot+[dashed,mark=none,
					samples=10,color=red,domain=0:42711104] 
					{1.192671e-05*x -1.727451e+01};
		
		\addlegendentry{Native, linear, $R^2= 0.996$}

		\end{axis}
		\end{tikzpicture}
		\qquad		
		\begin{tikzpicture}[scale=0.5]
		\begin{axis}[
		axis x line=middle,
		axis y line=middle,
		enlarge y limits=true,
		ylabel= CPU Time (sec),
		xlabel= {Number of Activities},
		x label style={at={(axis description cs:0.5,-0.08)},anchor=north},
		y label style={at={(axis description cs:-0.15,.5)},rotate=90,anchor=south},
		legend style={at={(0.5,1)}, anchor=north}
		]

		\addplot[only marks, blue] table  {result/abs-trad-act.dat};
		
		
		\addplot[only marks, green] table  {result/abs-interface-act.dat};
		
		
		\addplot[only marks, red] table  {result/abs-native-act.dat};
		
		
		\addplot+[dashed,mark=none,
					samples=10,color=green,domain=0:3840] 
					{-0.0005*x + 58.009};
		
		\addlegendentry{SQL Interface, constant}	
		
		\addplot+[dashed,mark=none,
					samples=10,color=red,domain=0:3840] 
					{0.0017*x + 28.105};
		
		\addlegendentry{Native, constant}	
		
		\end{axis}
		\end{tikzpicture}
		
		\caption{The comparison of abstraction phase}
		\label{fig:abstraction}
	\end{center}
\end{figure}

Figure \ref{fig:abstraction} shows the time for the abstraction phase of the four approaches. On the left, the time is shown as a function of the number of events in the log and on the right as a function of the number of activities. 

As expected, the time complexity of the native, SQL interface, and traditional approaches is linear because events in the log are already sorted. 
However, we cannot see the last dot of the traditional approach.
This is because the traditional approach cannot handle the biggest log containing 42M events due to out of memory exception.
Furthermore, the execution of the nested query leads to third order polynomial time complexity. 
It is considerably higher than the other approaches presented. 

As shown on the right hand side of Figure~\ref{fig:abstraction}, the number of activities does not affect the performance of the database approaches\footnote{due to the fact that the nested query is so time-consuming, we did not include it in some of the tests.}. However, for the traditional approach, there is an influence. This is due to the fact that, when scanning the log, at some points the internal data structures need to grow to accommodate for previously unseen activities. For $< 1000$ labels, the time is consistently around 30 seconds. For 1920 activities, the time is around 70 seconds and for 3840 activities, the time grows to 409 seconds. If the number of activities is known upfront, the implementation could try to allocate sufficient memory upfront, thus eliminating this effect. 

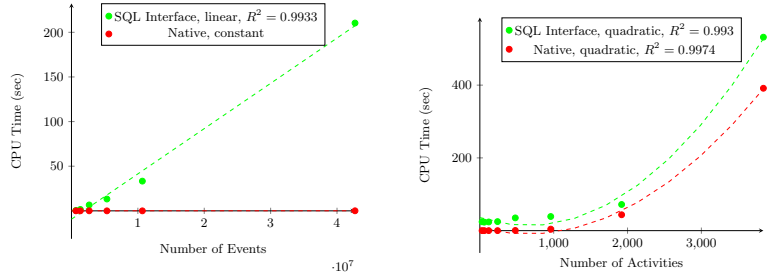
\begin{figure}[!t]
	\begin{center}
		
		\begin{tikzpicture}[scale=0.55]
		\begin{axis}[
		axis x line=middle,
		axis y line=middle,
		enlarge y limits=true,
		ylabel= CPU Time (sec),
		xlabel= {Number of Events},
		x label style={at={(axis description cs:0.5,0)},anchor=north},
		y label style={at={(axis description cs:-0.15,.5)},rotate=90,anchor=south},
		legend style={at={(0.5,1)}, anchor=north}
		]        
			
%

		\addplot[only marks, green] table  {result/ret-interface-event.dat};
		
		
		\addplot[only marks, red] table  {result/ret-native-event.dat};
		
		
%
		
		\addplot+[dashed,mark=none,
					samples=10,color=green,domain=0:42711104] 
					{5.072637e-06*x -9.336398e+00};
					
		\addlegendentry{SQL Interface, linear, $R^2 = 0.9933$}	
		
		\addplot+[dashed,mark=none,
					samples=10,color=red,domain=0:42711104] 
					{0.082906729};
		
		\addlegendentry{Native, constant}
	
		\end{axis}
		\end{tikzpicture}
		\qquad		
		\begin{tikzpicture}[scale=0.55]
		\begin{axis}[
		axis x line=middle,
		axis y line=middle,
		enlarge y limits=true,
		ylabel= CPU Time (sec),
		xlabel= {Number of Activities},
		x label style={at={(axis description cs:0.5,0)},anchor=north},
		y label style={at={(axis description cs:-0.15,.5)},rotate=90,anchor=south},
		legend style={at={(0.5,1)}, anchor=north}
		]        
				
		\addplot[only marks, green] table  {result/ret-interface-act.dat};
		
	
		\addplot[only marks, red] table  {result/ret-native-act.dat};
		
		
		\addplot+[dashed,mark=none,
					samples=10,color=green,domain=0:3840] 
					{5.075158e-05*x*x -6.682398e-02*x + 3.593839e+01};
					
		\addlegendentry{SQL Interface, quadratic, $R^2 = 0.993$}	
		
		\addplot+[dashed,mark=none,
					samples=10,color=red,domain=0:3840] 
					{3.905215e-05*x*x -5.041957e-02*x + 7.238121e+00};
					
		\addlegendentry{Native, quadratic, $R^2 = 0.9974$}	
		
		\end{axis}
		\end{tikzpicture}
		
		\caption{The comparison of retrieval phase}
		\label{fig:retrieval}
	\end{center}
\end{figure}

For the database approaches, the pre-processed data needs to be retrieved from the database into the process mining tool. Figure \ref{fig:retrieval} demonstrates the time needed for retrieval in the database approaches.
As shown in the figure, the retrieval phase in SQL interface is linear in the number of events, since each event is transferred. In contrast, the line of native is flat, since the events themselves are not retrieved, but only the non-zero elements in the directly follows relation.
This is shown by the right-hand figure, showing the influence of the number of activities in retrieval phase. Both native and SQL interface demonstrate a second order polynomial time complexity since retrieving the DFR is (worst case) quadratic in the number of activities.

\begin{figure}[!t]
	\centering
	\begin{minipage}{.5\textwidth}
		\centering
			\begin{center}
		
		\begin{tikzpicture}[scale=0.6]
		\begin{axis}[
		axis x line=middle,
		axis y line=middle,
		enlarge y limits=true,
		ylabel= CPU Time (sec),
		xlabel= {Number of Activities},
		x label style={at={(axis description cs:0.5,0)},anchor=north},
		y label style={at={(axis description
				cs:-0.15,.5)},rotate=90,anchor=south},
		legend style={at={(0.5,1)}, anchor=north}
		]

		\addplot[only marks, blue] table  {result/min-trad-act.dat};
		
		\addplot[only marks, orange] table  {result/min-nested-act.dat};
				
		\addplot[only marks, green] table  {result/min-interface-act.dat};
		
		\addplot[only marks, red] table  {result/min-native-act.dat};
	
		\addplot+[dashed,mark=none,
					samples=100,color=blue,domain=0:3840] 
					{1.332704e-09*x*x*x + 6.678386e-06*x*x -3.502385e-03*x + 3.039503e-01};
					
	 	\addlegendentry{Traditional, cubic, $R^2= 1$}	
		
		\addplot+[dashed,mark=none,
					samples=10,color=orange,domain=0:801] 
					{1e-10*x*x*x - 2e-07*x*x + 0.0001*x + 0.021};
		
		\addlegendentry{Nested SQL, cubic, $R^2= 1$}	
			
		\addplot+[dashed,mark=none,
					samples=100,color=green,domain=0:3840] 
					{2.841381e-09*x*x*x + 1.790142e-06*x*x -7.670717e-04*x + 1.055664e-01};
				
		\addlegendentry{SQL Interface, cubic, $R^2= 1$}	
			
		\addplot+[dashed,mark=none,
					samples=10,color=red,domain=0:3840] 
					{1e-09*x*x*x + 7e-06*x*x - 0.0038*x + 0.357};
					
		\addlegendentry{Native, cubic, $R^2= 1$}	
				
		\end{axis}
		\end{tikzpicture}
		
		\caption{The comparison of mining\\phase}
		\label{fig:mining}
	\end{center}
	\end{minipage}%
	\begin{minipage}{.5\textwidth}
		\centering
			\begin{center}
		
		\begin{tikzpicture}[scale=0.6]
		\begin{axis}[
		axis x line=middle,
		axis y line=middle,
		enlarge y limits=true,
		ylabel= CPU Time (sec),
		xlabel= {Number of Events},
		x label style={at={(axis description cs:0.5,0)},anchor=north},
		y label style={at={(axis description
				cs:-0.15,.5)},rotate=90,anchor=south},
		legend style={at={(0.5,1)}, anchor=north}
		]

		\addplot[only marks, orange] table  {result/abs-nested-noindex-event.dat};
		
		\addplot[only marks, black] table  {result/abs-nested-index-event.dat};
		
		\addplot[only marks, purple] table  {result/abs-native-index-event.dat};
	
		\addplot+[dashed,mark=none,
					samples=100,color=orange,domain=0:27362] 
					{-2.595230e-11*x*x*x + 2.303960e-06*x*x -4.331604e-03*x + 6.498454e+00};
					
	 	\addlegendentry{Nested without index, cubic, $R^2= 1$}	
		
		\addplot+[dashed,mark=none,
					samples=10,color=black,domain=0:3502336] 
					{-2.8991032426 + 0.0000421135*x*ln(x)/ln(10)};
		
		\addlegendentry{Nested with index, $x \cdot \log{}x$, $R^2= 0.9987$}	
			
		\addplot+[dashed,mark=none,
					samples=100,color=purple,domain=0:3502336] 
					{8.551608e-06*x -2.149149e-01};
				
		\addlegendentry{Native with index, linear, $R^2= 0.9903$}	
				
		\end{axis}
		\end{tikzpicture}
		
		\caption{The comparison of abstraction phase with index for native vs nested SQL}
		\label{fig:abs-index}
	\end{center}
	\end{minipage}
\end{figure}

The time complexity of Inductive Miner is worst-case cubic in the number of activities, which is clearly shown in Figure \ref{fig:mining}. There is no difference between the four approaches since they use the same implementation of the Inductive Miner.

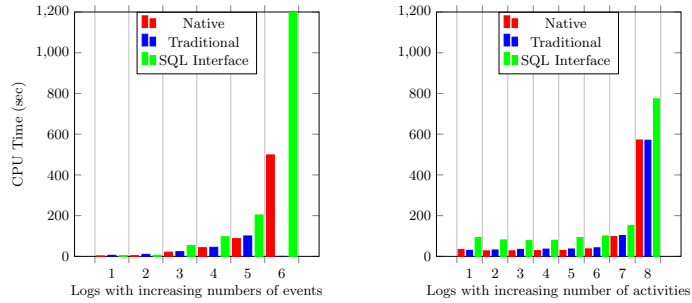
\begin{figure}[!t]
	\begin{center}
		\begin{tikzpicture}[scale=0.6]
\begin{axis}[
		height=7cm,
		width=7cm,
	xlabel = Logs with increasing numbers of events,
	x tick label style={
		/pgf/number format/1000 sep=},
	ylabel=CPU Time (sec),
	ymin=0, ymax=1200,
	legend style={at={(0.5,1)}, anchor=north},
	ybar interval=0.7,
]
\addplot[color=red,fill] 
	coordinates {(1,1.76528562)	
		(2,2.931695098)	
		(3,20.83167523)	
		(4,42.65929242)	
		(5,87.19083514)
		(6,498.0135499)
		(7,0)};
	
\addplot[color=blue,fill] 
	coordinates {(1,5.186868364)
		(2,9.932372979)
		(3,23.18000195)
		(4,44.54346845)
		(5,100.3079988)
		(6,0)
		(7,0)};
	
\addplot[color=green,fill] 
	coordinates {(1,3.179151487)	
		(2,5.783350023)	
		(3,53.12093149)	
		(4,96.29756197)	
		(5,203.3181072)	
		(6,1256.199314)	
		(7,0)};
				
\legend{Native, Traditional, SQL Interface}
\end{axis}
\end{tikzpicture}		
		\qquad	
		\begin{tikzpicture}[scale=0.6]
\begin{axis}[
		height=7cm,
		width=7cm,
	xlabel = Logs with increasing number of activities,
	x tick label style={
		/pgf/number format/1000 sep=},
	ymin=0, ymax=1200,
	legend style={at={(0.5,1)}, anchor=north},
	ybar interval=0.7,
]
\addplot[color=red,fill] 
	coordinates {(1,33.985949)	(2,26.99746278)	(3,27.06632475)	(4,28.124486)	(5,29.08949482)	(6,36.68979301)	(7,97.67376409)	(8,571.4818714) (9,0)};
	
\addplot[color=blue,fill] 
	coordinates {(1,29.65187966)	(2,31.42069137)	(3,34.4231725)	(4,35.72406844)	(5,36.12881169)	(6,42.10187488)	(7,102.6489599)	(8,570.4220423) (9,0) };
	
\addplot[color=green,fill] 
	coordinates {(1,92.91450251)	(2,80.82210946)	(3,77.1993732)	(4,78.88910477)	(5,92.28956149)	(6,100.4323352)	(7,152.1508656)	(8,773.8644792) (9,0) };
				
\legend{Native, Traditional, SQL Interface}
\end{axis}
\end{tikzpicture}	
		
		\caption{The comparison of all phases for native vs traditional and SQL interface}
		\label{fig:barchart}
	\end{center}
\end{figure}

Figure \ref{fig:barchart} denotes the total time of Figure \ref{fig:abstraction}-\ref{fig:mining}, i.e. the total time of abstraction, retrieval, and mining phases.
In Figure \ref{fig:barchart}, we do not include the nested query because it is so time-consuming (the experiments with nested query were stopped until the number of events reached 14K).
However, we still tried to improve the performance of nested query by adding indices in case and timestamp columns.
Even though the cubic complexity in the nested query is reduced to $x \cdot \log{}x$ (with $x$ is the total number of events), the native approach still outperforms the nested query because the former has linear complexity as shown in Figure \ref{fig:abs-index}.
Note that the linearithmic comes from the fact that H2 database uses B-tree index, hence finding an element is $\mathcal{O}(\log{}x)$.
There are $x$ rows for which we need to perform this look up, therefore the complexity is $\mathcal{O}(x \cdot \log{}x)$.


From Figure \ref{fig:abs-index} and \ref{fig:barchart} it becomes clear that for large numbers of events and/or large numbers of activities, the approach using native operator outperforms the other database approaches. 
Both native and traditional approaches show relatively similar performance, except for the log with the biggest number of events (the 6$^{th}$ log in the left chart of Figure \ref{fig:barchart}).
The 6$^{th}$ bar of the traditional approach does not exist because the approach cannot handle the log anymore.
This is to be expected as this approach works in memory. 
As data grows, the event logs no longer fit into memory and then the database approaches are required. 
There is however another important motivation for in-database processing.

In most information systems, data is stored in relational databases. To do process mining, this data needs to be exported in the form of an XES file, which required sorting of the events on timestamps, i.e. this preparatory phase has the same time complexity as in-database abstraction.
In the next experiment, we take into account two additional steps in the traditional approach, namely the exporting phase (i.e. exporting event data from database to a Comma Separated Value (CSV) form) and the conversion phase (i.e. conversion from the CSV file to an XES event log file).
These two phases are not relevant to database approaches since they directly access the data source.

In Table \ref{tab:exp} we show the complete chain of discovery using traditional technique, starting from exporting until mining.
We can see from this table that, for example, there is additional 10 seconds to export and convert event data with 100K traces and 660K events.
This additional time increases up to 466 seconds for event data with 500K traces and 13M events.

From Table \ref{tab:exp}, we deduce that the export (11.3 \%) and conversion (16.9 \%) phases take up to 28.2\% of the whole process, while the rest is dominated by the abstraction phase (71.8 \%). The actual mining only takes 0.1 \% of the time. 
This shows that there is 28.2 \% overhead in the traditional approach.
A direct process mining approach to database, as proposed by the native operator, is an excellent solution to remove this overhead.

\begin{table}[!bt]	
	\centering
	\footnotesize
	\caption{Computation time in different phases in traditional process discovery (measured in seconds)}\scriptsize
	\begin{tabular}{|l|S[table-format=8.0]|S[table-format=3.0]|S[table-format=3.3]|S[table-format=3.6]|S[table-format=3.6]|S[table-format=1.9]|S[table-format=3.6]|}
		\hline
		Traces & {Events} & {Activities} & {Exporting} & {Conversion} & {Abstraction} & {Mining} & {Exporting+Conversion} \\
		\hline
		100000 & 667361 & 30 & 2.532 & 7.784738 & 43.360468 & 0.097992955 & 10.316738 \\
		100000 & 1352645 & 50 & 5.505 & 16.717697 & 87.035568 & 0.028197078 & 22.222697 \\
		100000 & 2718451 & 200 & 30.871 & 31.705399 & 185.524574 & 0.140519505 & 62.576399 \\
		100000 & 4039403 & 400 & 55.275 & 74.437021 & 260.539442 & 0.076435603 & 129.712021 \\
		100000 & 4453320 & 100 & 68.872 & 78.903779 & 275.585310 & 0.195748059 & 147.775779 \\
		500000 & 3346890 & 30 & 39.525 & 53.646275 & 250.444424 & 0.005964106 & 93.171275 \\
		500000 & 6761572 & 50 & 90.275 & 133.006643 & 462.305078 & 0.128949184 & 223.281643 \\
		500000 & 13629325 & 200 & 182.267 & 284.626551 & 950.972794 & 0.136748262 & 466.893551 \\
		\hline
	\end{tabular}
	\label{tab:exp}
\end{table}


\section{Conclusion}
\label{sec:concl}

In this paper we focus on direct process discovery on relational databases.
We propose a native SQL operator for computing one of the intermediate structures, namely the Directly Follows Relation (DFR).
Using this operator, users can easily express their query and get DFR values from the underlying data. 
Moreover, the operator does not contain any nested form which causes bad performance. 

The native operator has been implemented in H2 database and we provide a specialized implementation of the state-of-the-art process mining technology to show applicability of the work.

\bibliographystyle{plain}
\bibliography{main-bib}
	
\end{document}